\documentclass[letterpaper,journal]{IEEEtran}
\usepackage{amsmath,amssymb,amsfonts}
\usepackage{graphicx}
\usepackage{booktabs}
\usepackage{hyperref}

\begin{document}
 
\twocolumn[
\begin{@twocolumnfalse}
\Huge {IEEE copyright notice} \\ \\
\large {\copyright\ 2023 IEEE. Personal use of this material is permitted. Permission from IEEE must be obtained for all other uses, in any current or future media, including reprinting/republishing this material for advertising or promotional purposes, creating new collective works, for resale or redistribution to servers or lists, or reuse of any copyrighted component of this work in other works.} \\ \\
	
{\Large Published in \emph{IEEE Embedded Systems Letters} (Volume: 16, Issue: 1, March 2024, pp. 25-28)} \\ \\
{\Large DOI: \href{https://doi.org/10.1109/LES.2023.3240625}{10.1109/LES.2023.3240625}} \\ \\

\end{@twocolumnfalse}
]

\title{Evaluating the effects of reducing voltage margins for energy-efficient operation of MPSoCs}
\author{
Diego~V.~Cirilo~do~Nascimento, Kyriakos~Georgiou, Kerstin~I.~Eder and Samuel~Xavier-de-Souza
\thanks{D. V. C. do Nascimento is with Instituto Federal do Rio Grande do Norte - IFRN, Natal-RN, Brazil.

K. Georgiou and K. I. Eder are with the University of Bristol, UK.

S. Xavier-de-Souza is with Universidade Federal do Rio Grande do Norte - UFRN, Natal-RN, Brazil.
}
}

\maketitle

\begin{abstract}
  Voltage margins, or guardbands, are imposed on DVFS systems to account for process, voltage, and temperature variability effects.
  While necessary to assure correctness, guardbands reduce energy efficiency, a crucial requirement for embedded systems.
The literature shows that error detection techniques can be used to maintain the system's reliability while reducing or eliminating the guardbands.
  This letter assesses the practically available margins of a commercial RISC-V MPSoC while violating its guardband limits.
  The primary motivation of this work is to support the development of an efficient system leveraging the redundancy of multicore architectures for an error detection and correction scheme capable of mitigating the errors caused by aggressive voltage margin reduction.
  For an equivalent performance, we achieved up to 27\% energy reduction while violating the manufacturer's defined guardband, leaving reasonable energy margins for further development.
\end{abstract}


\section{Introduction}

\IEEEPARstart{E}{nergy} efficiency is crucial for mobile and embedded systems, where battery size, autonomy, and power dissipation are under hard constraints.
Such systems must comply with these limitations while being capable of delivering the high performance needed by demanding Edge Computing and IoT devices \cite{georgiou2018iot}.

The supply voltage significantly impacts the total power consumption, hence should be kept at a minimum.
Signal propagation delays, however, are inversely proportional to the operating voltage and are a limiting factor for the maximum clock frequency~\cite{butts2000static}.

Dynamic voltage and frequency scaling (DVFS) techniques have been developed as means to meet processing requirements at minimal power dissipation~\cite{burd2000dynamic}.
These operating voltage/frequency points are defined \textit{a priori} and must comply with critical-path timing limitations~\cite{xia2017voltage}.
The critical path delays are affected by process, voltage, and temperature (PVT) variability, so safety voltage margins or ``guardbands'' are imposed on the DVFS system, assuring a reliable operation.
Critical path violation errors may occur if the system is otherwise allowed to run too close to the physical limits.
Ongoing miniaturization increases variability during chip manufacture, and these guardbands are vital for lower technology nodes.
Consequently, the addition of voltage guardbands trades energy efficiency for correctness~\cite{das2009addressing}.

Seeking aggressive guardband reduction, error-detection-based circuits, such as Razor~\cite{blaauw2008razor}, use special delayed-path flip-flops to detect and correct timing failures.
These failures are then reported and used for adaptive voltage scaling.
The main advantage is that the voltage guardbands are effectively removed at the expense of deep circuit modifications, which add design and verification overheads, limiting its application in existing designs.

Bacha and Teodorescu~\cite{bacha2013dynamic} proposed a less intrusive architectural approach, leveraging the existing Error Detection and Correction (EDAC) hardware in Intel Itanium processors for voltage margin elimination.
The system is allowed to run in a configuration susceptible to timing errors but is kept functional by relying on the EDAC features.
By keeping the system at an operating point where the error rate is low enough, the energy gains surpass the recovery costs with a negligible performance penalty.

Simevski et al.~\cite{simevski2020pisa} and Silva et al.~\cite{silva2020cevero} have proposed on-demand EDAC functionalities based on core redundancy of Multi-processor System-on-Chip (MPSoC), where cores in the architecture can be arranged in lockstep execution for dual or triple redundancy when needed. To the best of our knowledge, this type of EDAC has not been used for guardband reductions. The advantage comes from the possibility of switching off EDAC and using the redundant cores for regular computations in non-critical code or non-critical environments. 

In the interest of assessing the viability of using MPSoC core-redundancy-based EDAC schemes for voltage guardband reduction, this letter presents characterization experiments of a commercial MPSoC, the Greenwaves Technologies GAP8.
This characterization consists of violating the published manufacturer's voltage guardbands by overclocking the chip and measuring the available energy margins, practical voltage/frequency limits, and error characteristics.

Our experiments achieved frequencies up to $2.5\times$ higher than those specified in the datasheet before any errors or lockups were detected.
When comparing performance-equivalent operating points, we were able to reduce the energy by 27\%, confirming that margins are available for implementing an error detection and correction scheme to ensure reliability and energy efficiency in these extreme settings.

In the rest of the text, Section \ref{sec:literature} presents related works and supporting literature data, Section \ref{sec:methodology} presents the experimental methodology, Section \ref{sec:results} details the results, and Section \ref{sec:conclusion} contains our conclusions.

\section{Related Works}
\label{sec:literature}

Two main topics are of interest to this letter, aggressive voltage margin elimination and error detection methods in multiprocessor chips.

\subsection{Voltage margins elimination}
Bacha and Teodorescu~\cite{bacha2013dynamic} used a firmware-based solution on an Intel Itanium 9560, aggressively reducing operating voltage until the on-chip EDAC system started reporting recoverable errors. 
Actively controlling the voltage to keep the error rate between set boundaries,
they achieved an average reduction of 20\% in power consumption, with an error-recovery time overhead of less than 1\%.
In a subsequent paper, a more efficient approach, with dedicated hardware targeting sensitive cache lines, achieved a power reduction of 33\%~\cite{bacha2014using}.

Leng et al.~\cite{leng2021predictive} reduced voltage margins in NVIDIA GPUs by predicting the $V_{\rm min}$, or minimal error-free voltage.
Based on the observation that this $V_{\rm min}$ primarily depends on voltage droops caused by software operation, i.e., the available margins are application dependent, they derived a prediction scheme based on the analysis of performance counters.
They found that voltage can be reduced by up to 20\% without errors.

Papadimitriou et al.~\cite{papadimitriou2017voltage,papadimitriou2017harnessing} exploited the available margins at multiple platforms in a series of papers, using available EDAC hardware error reporting as a guideline.
They ran benchmarks from the SPEC2600 suite and collected information about multiple margins, from error-free operation up to system lockup.
By reducing the voltage margins, they achieved power savings of 18\% in ARMv8 processors, 20\% in Intel processors, 25\% on NVIDIA Fermi and Kepler GPUs, and up to 90\% for FPGA on-chip memories~\cite{papadimitriou2020exceeding}.

In our review, no equivalent tests were performed on MPSoCs, probably due to the lack of EDAC features in devices readily available on the market.

\subsection{Error detection in MPSoCs}

Simevski et al.~\cite{simevski2020pisa} investigated the usage of idle cores in multiprocessors for N-modular redundancy schemes. Subsequently, they developed the Waterbear framework, a fault-tolerant multicore architecture with switchable operating modes, using parallelism or redundancy-based fault-tolerance on demand.

Silva et al.~\cite{silva2020cevero} presented the CEVERO architecture, a fault-tolerant MPSoC based on the open-source PULP Platform~\cite{pullini2018mr}.
This architecture uses a switchable Dual-Modular Redundancy (DMR) scheme aided by ``safe'' status registers, which can return the cores to the last known-good state and resume operation.

Rogenmoser et al.~\cite{rogenmoser2022ondemand} developed an On-Demand Redundancy Grouping (ODRG) scheme for the PULP platform, where cores in a cluster can be arranged either for parallel computing or fault-tolerant mode, running a Triple Modular Redundancy (TMR) block with three cores in lockstep.
Their approach uses a voter, maintaining execution in the event of a mismatch in the output of one of the cores and later synchronizing the faulty core.

Both Waterbear and ODRG approaches need three cores for error recovery. In comparison, the CEVERO architecture uses two cores and extra state registers, saving on area and power consumption at the cost of a minor performance penalty during the recovery process.

Our present work aims to assess the available energy margins when exceeding guardband parameters in an architecture similar to the one used by Silva et al., supporting the  development of a guardband reduction methodology based on the CEVERO architecture.

\section{Experimental Methodology}
\label{sec:methodology}

Using a PULP-based MPSoC, two experiments were arranged, one for detecting computation errors and lockups while exceeding the published voltage and frequency guardbands and the other using a real-world application to measure energy consumption in extreme settings.
The goal was to assess the available margins for these use cases, serving as supporting evidence for the future development of an error-detection-based methodology to reduce voltage margins.

\subsection{Hardware Setup}
The base hardware is the Greenwaves Technologies GAP8, a commercial implementation of the PULP's Mr. Wolf~\cite{pullini2018mr} chip.
The main benefits of using this architecture are the open nature of most of the design, facilitating further development, modularity, and state-of-the-art energy efficiency features.

It is composed of a Fabric Controller (FC) core and an eight-core cluster.
The FC and Cluster are on separate clock domains, which can be controlled by an on-chip Frequency Locked Loop (FLL) device.
The system also has an onboard DC-DC converter able to supply voltages from $1.0\ V$ up to $1.2\ V$ in $50\ mV$ steps.
Table~\ref{tab:limits} shows the datasheet guardbands.

\begin{table}
  \centering
  \caption{GAP8 Guard band values~\cite{greenwavestechnologiessas2018gap8}.}
  \label{tab:limits}
  \begin{tabular}{@{}lrr@{}}
    \toprule
    Voltage   & $FC_{\rm max}$ & $Cluster_{\rm max}$ \\ \midrule
    1200 mV   & 250 MHz    & 170 MHz       \\
    1150 mV   & 225 MHz    & 149 MHz       \\
    1100 mV   & 200 MHz    & 129 MHz       \\
    1050 mV   & 175 MHz    & 108 MHz       \\
    1000 mV   & 150 MHz    & 87 MHz       
  \end{tabular}
\end{table}

A GAP8 development board, called \emph{Gapuino}, is used to provide hardware support, external power supply and communications, and test points for voltage and current measurements.

\subsection{Measurements}

A GPIO port is used as a trigger to start the time and power measurement. 
The measurements are performed using a MAGEEC Power Measurement board\cite{mageec2016power}, based on ST Discovery STM32F407VG.
The power is measured by a shunt resistor available at the GAP8 power supply path, isolated from the rest of the board but including the on-chip DC-DC converter.
The energy of the operation is calculated using execution time and average power.

\subsection{Experiment A setup}%
\label{sec:Test setup}

\begin{figure}
\begin{center}
  \includegraphics[width=\linewidth]{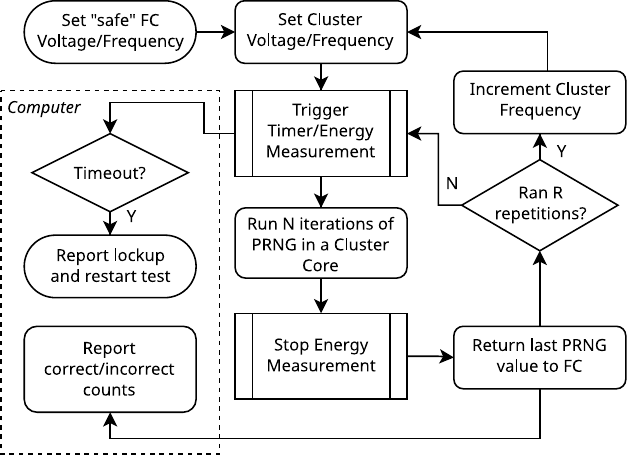}
\end{center}
\caption{Test flowchart.}
\label{fig:flowchart}
\end{figure}

The first experiment aims to discover the operating limits of the chip. As no hardware EDAC features are present on-chip, we used a pseudorandom number generator (PRNG) producing known sequences of $N$ numbers in variable frequency/voltage settings, which could be later verified for correctness.
A flowchart of the test can be seen in Fig.~\ref{fig:flowchart}.

We proposed this workload due to its susceptibility to data corruption, deviating the output sequence in the event of any error in the calculations.
To counter the lack of hardware error-detection capabilities, periodic checking of the results was necessary, comparing the last value of the generated sequence with known-good values.
Any mismatch in these values results from calculation or memory errors during the execution.

The test control happens in the FC core, which is set as a ``safe'' core, and the PRNG is run in a Cluster core, where the clock can be set to extreme values independently.

After each $R$ run, the last value is returned to be compared to the reference.
If different, the test is flagged as an error.
The test is flagged as a lockup if no response is received after a timeout.

The tests were run for all available voltages and frequencies starting at $200\ MHz$, increasing in $2\ MHz$ steps until the chip stopped responding.
The starting frequency was decided empirically in preliminary tests, where errors only started to occur beyond this frequency, even for the lowest voltage.

For any given frequency/voltage pair, the test was run with increasing problem sizes, from 50 thousand to 1 million numbers in 50 thousand steps, to check if error probability was related to runtime.
The test duration depended on the problem size and set frequency, ranging from approximately $35\ ms$ to 32 seconds.

A total number of 678,440 tests were executed.

\subsection{Experiment B setup}

The second experiment uses the same hardware setup as Experiment A, and a real-world application as a workload, aiming to collect data on the energy gains achievable by running the system in extreme settings.

The application is a satellite communications algorithm developed by De Lima \textit{et al.} \cite{delima2021parallel} for the GAP8 platform.
This algorithm runs in parallel, using all eight cores of the cluster, and performs multiple operations, including data movement and integer FFT.

The tests ran in all available voltage levels and frequencies ranging from $80\ MHz$, inside the guardband for all voltages, to $200\ MHz$, outside the guardband for all voltages. These settings were defined based on the maximum frequencies reached at $1\ V$ in Experiment A.

\section{Results}
\label{sec:results}

\begin{figure}
\begin{center}
  \includegraphics[width=\linewidth]{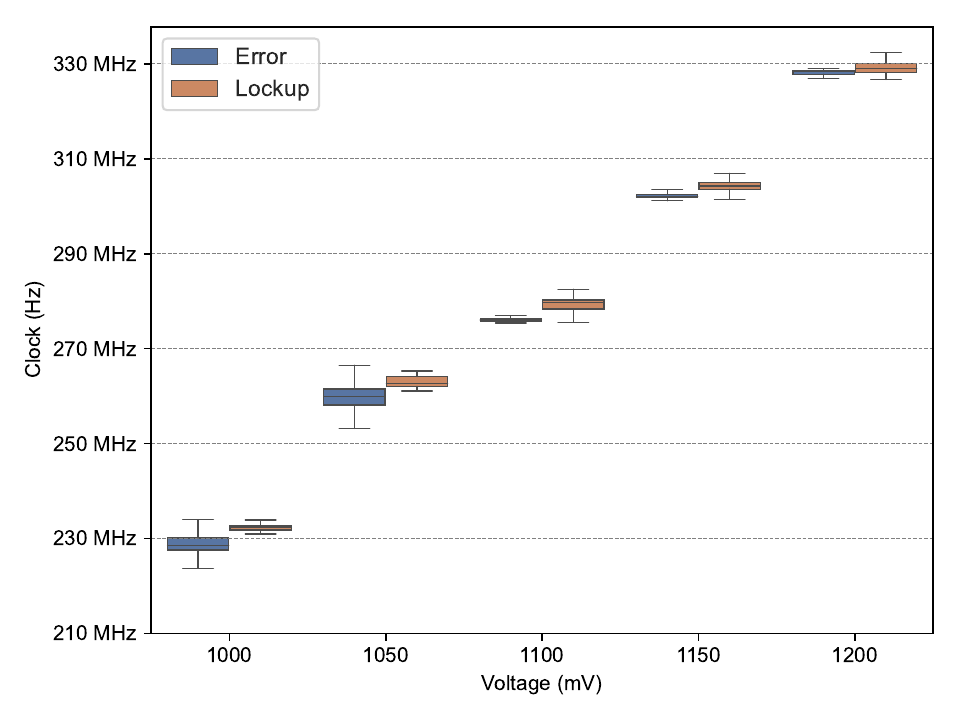}
\end{center}
\caption{Error and lockup distribution.}
\label{fig:boxplot}
\end{figure}

Fig.~\ref{fig:boxplot} presents the error and lockup figures of the whole dataset of Experiment A, with a guardband reference line at the bottom of the graph.
It is possible to observe that, apart from a few outliers, the failures occur in a narrow band, without much margin between the point where the error rate starts to rise and the total loss of response.
The error probability did not grow with the problem size, being primarily dependent on the critical operating points.
Hence the narrow failure points distribution shown in Fig.~\ref{fig:boxplot}.
As the voltage increases, the lockups started to occur before the errors are detected.
For lower voltages, the data shows a tendency that the errors start to happen at even lower voltages than the lockups, probably due to the critical path violations being the primary source of errors.

In any case, from the guardband to the minimum value, we have at least two times the clock frequency, reaching 2.5$\times$ for the lowest tested voltage ($1\ V$).

Being able to reach higher frequencies for the same voltage can be translated to reduced energy consumption as execution times decrease. A caveat in this approach is the temperature increase in the components, which could be handled by treating temperature violations as errors.
With the proper hardware setup, further voltage reduction could be exploited while respecting applications' timing requirements. Core-redundancy-based EDAC methodologies could mitigate the errors and lockups, and the reported error rate used to control an active voltage scaling scheme.

\begin{figure}
\begin{center}
  \includegraphics[width=\linewidth]{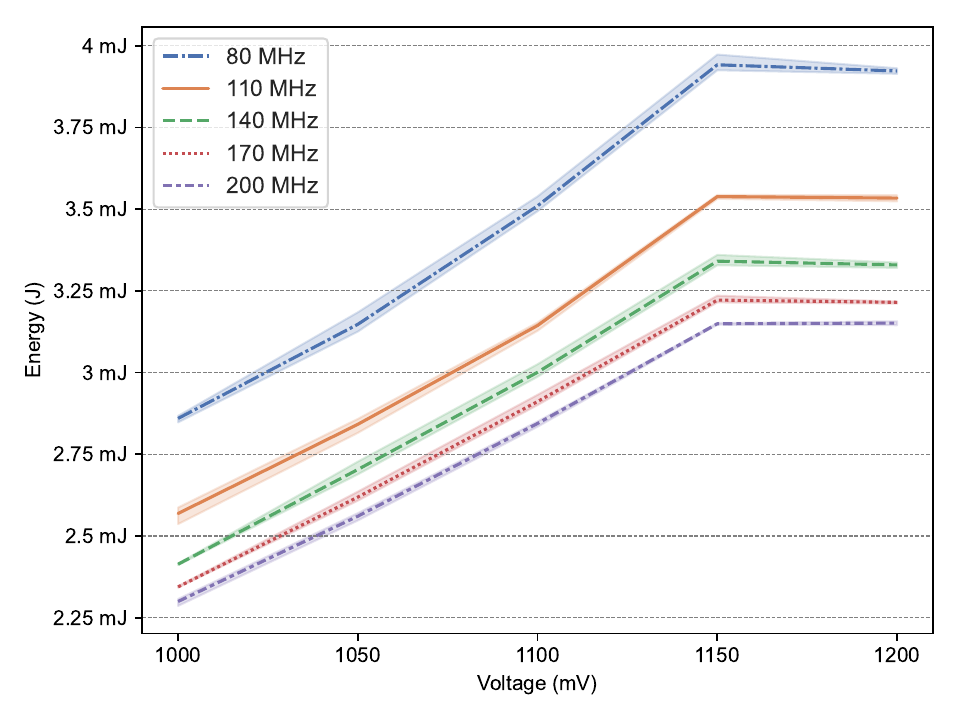}
\end{center}
\caption{Energy consumption of SBCDA decoder.}
\label{fig:sbcda-isoperf}
\end{figure}

For Experiment B, Fig.~\ref{fig:sbcda-isoperf} presents the energy consumption characteristic of the circuit for a given performance.
Up to 27\% of energy can be saved by overclocking the system, as the time reduction compensates for the power increase.

The flattening of the energy consumption between the $1.15\ V$ and $1.2\ V$ settings is probably due to power circuitry effects.
We are not able to measure the core voltages, but this voltage change can be confirmed by the increased maximum ``error-free'' frequency.

\section{Conclusion}
\label{sec:conclusion}
This letter presented experiments characterizing the GAP8 MPSoC when violating the manufacturer-defined voltage guardband, providing evidence and motivation for developing novel guardband reduction techniques.

For multicore architectures, the inherent redundancy of cores can be exploited in a fault-tolerant arrangement, which in turn can be used to warrant correct operation in guardband-violating settings.
These extreme settings can improve energy efficiency, for example, in applications unfit for parallelization.

The energy reduction of about 27\% achieved in the experiments provides a target consumption for the complete system. We can expect more significant gains if we consider that the maximum energy reduction was in a ``safe'' configuration for the energy measurement experiment as no fault-tolerant system was employed. Therefore, considering the failure threshold shown in Experiment A, there are still available margins to be harvested. Additionally, due to the limitations of the on-chip DC-DC converter, it was not possible to go lower than $1\ V$. Fig. \ref{fig:sbcda-isoperf} shows a trend of higher energy efficiency towards lower voltages. Further energy reductions could be possible by allowing errors in suitable applications, such as image processing or signal processing algorithms, in an approximate computing approach.

\bibliographystyle{IEEEtran}
\bibliography{ref.bib}

\end{document}